# Fabrication of Spin-1/2 Heisenberg Antiferromagnetic Chains via Combined On-surface Synthesis and Reduction for Spinon Detection

Xuelei Su[1,†], Zhihao Ding[1,†], Ye Hong[1,†], Nan Ke[1], KaKing Yan[1,*], Can Li[2,3,*], Yifan Jiang[1,*] and Ping Yu[1,*]

**Spin-1/2 Heisenberg antiferromagnetic chains are excellent one-dimensional platforms for exploring quantum magnetic states and quasiparticle fractionalization. Understanding its quantum magnetism and quasiparticle excitation at the atomic scale is crucial for manipulating the quantum spin systems. Here, we report the fabrication of spin-1/2 Heisenberg chains through on-surface synthesis and in-situ reduction. A closed-shell nanographene is employed as a precursor for Ullman coupling to avoid radical fusing, thus obtaining oligomer chains. Following exposure to atomic hydrogen and tip manipulation, closed-shell polymers are transformed into spin-1/2 chains with controlled lengths by reducing the ketone groups and subsequent hydrogen desorption. The spin excitation gaps are found to decrease in power-law as the chain lengths, suggesting its gapless feature. More interestingly, the spinon dispersion is extracted from the inelastic spectroscopic spectra, agreeing well with the calculations. Our results demonstrate the great potential of fabricating desired quantum systems through a combined on-surface synthesis and reduction approach.**

[1]School of Physical Science and Technology, ShanghaiTech University, 201210 Shanghai, China

[2]Key Laboratory of Artificial Structures and Quantum Control (Ministry of Education), Shenyang National Laboratory for Materials Science, School of Physics and Astronomy, Shanghai Jiao Tong University, Shanghai 200240, China

[3]Tsung-Dao Lee Institute, Shanghai Jiao Tong University, Shanghai, 200240, China

In one-dimensional antiferromagnetic systems, the classical Néel spin order is suppressed by enhanced quantum spin fluctuations resulting from low dimensionality, thereby giving rise to exotic quantum phenomena. As a distinguished model, one-dimensional Heisenberg spin chains have garnered significant experimental and theoretical interest in exploring strongly correlated quantum many-body states and fractional excitation. In 1983, Haldane postulated that integer antiferromagnetic Heisenberg chains would demonstrate a spin excitation gap, known as the Haldane gap, between the singlet ground state and the first excited states[1]. Both theoretical[2-4] and experimental[5] investigations have further confirmed the presence of such topologically protected edge states at the termini of integer spin chains. In contrast, half-integer spin chains are anticipated to possess a gapless excitation spectrum for which a precise analytic solution, namely the Bethe ansatz, exists[6-8]. The ground state of half-integer spin chains represents a macroscopic singlet state, a manifestation of a quantum criticality denoted as a Tomonaga-Luttinger liquid[9]. Fractional quasiparticle excitations termed spinons as elemental excitations are produced in pairs within spin-1/2 Heisenberg antiferromagnetic chains, resulting in a continuum excitation spectrum from the ground state. Quasi S = 1/2 spin chains have been observed in inorganic oxide compounds using ensemble techniques such as neutron scattering or resonant inelastic X-ray scattering[10,11]. Although the atomically precise synthesis and characterization of spin-1/2 Heisenberg chains are of paramount importance for investigating and engineering the quantum states of many-body systems, a suitable platform for high spatial resolution detection is currently limited and highly desired.

The scanning tunneling microscope (STM) provides a great opportunity to precisely construct spin chains atom by atom, allowing for exploring electronic states and spin excitations at the atomic scale. Utilizing STM, spin chains and arrays have been precisely formed on surfaces using high spin atoms, unveiling intricate magnetic exchange interactions and topological properties[12]. Quantum magnets demonstrating resonating valence bond states and quantum phase transitions have been successfully realized using atom manipulation with finite sizes of spin-1/2 atoms[13,14] However, the magnetic anisotropy of transition metal ions and limited atom lengths have restricted the investigation of the quantum spin states and fractional excitation in the spin-1/2 Heisenberg chain. The emergence of on-surface synthesis [15-21] has led to the discovery of open-shell nanographene as promising building blocks for creating quantum spin systems[22-27]. Notably, one-dimensional spin-1 chains made of π-electron magnetic nanographene have unveiled the Haldane gap and topological edge states, highlighting the potential of combined on-surface synthesis and STM atomic detection[10,28] Another compelling example demonstrates the successful engineering of diverse topological phases in a nanographene spin-chain with alternating exchange interaction, enabling precise control over spin chain length, parities, and terminations[29]. Consequently, the spin-1/2 Heisenberg chain based on π-electron magnetism stands as an ideal platform for studying quantum many-body phenomena using on-surface synthesis and STM techniques to observe fractional spinon excitation with atomic precision. However, due to the requirements of obtaining long spin-1/2 Heisenberg chains, it has yet to be realized.[30,31]

Here, we report atomically precise synthesis and characterization of spin-1/2 Heisenberg chains through a tandem on-surface synthesis and *in-situ* reduction with STM capabilities for precise spin detection at the atomic scale. In our experiments, we use olympicenone[32] with bromo substitutions as a precursor for Ullman coupling, whose π-electron magnetism is quenched by the presence of carbonyl group. Employing closed-shell nanographene as a precursor for Ullman coupling avoids undesired side reactions such as radical-assisted coupling or random fusing between building blocks[33]. To recover the π-electron magnetism, the olympicenone oligomers are subsequently exposed to hydrogen atoms on the surface to facilitate a reductive deoxygenation reaction. Subsequently, one of the two hydrogens at the $sp^3$-$CH_2$ site can be dissociated through atom manipulation, producing delocalized π radical in the system and the formation of the desired spin-1/2 Heisenberg chains. The reaction products are characterized by non-contact atomic force microscopy (nc-AFM) and STM. The quantum collective excitations of spin-1/2 chains are comprehensively investigated by scanning tunneling spectroscopy (STS) measurement and Heisenberg model calculations. An even-odd parity effect is observed in spin-1/2 Heisenberg chains at short lengths, detecting spin excitations in even-numbered chains and Kondo resonance at the odd sites in odd-numbered spin-1/2 chains, respectively. The spin excitation gaps exhibit a power-law decay with the chain lengths, indicating its gapless feature. More crucially from a fundamental standpoint, fractional spinon excitation with $\frac{\pi}{2}J|\sin(\vec{a}\cdot\vec{q})|$ dispersion is detected at long chains, which is consistent with the Heisenberg model calculations.

**Results**

**On-surface synthesis of spin-1/2 chains.**

Fig. 1 shows the stepwise procedure for fabricating spin-1/2 Heisenberg chains based on olympicene building blocks. Olympicene is an aromatic hydrocarbon composed of five rings arranged in an Olympic rings-shape, resulting in an unpaired radical known as olympicenyl radical due to the inherent sublattice imbalance. Although olympicenyl radical has attracted significant interest as a critical precursor, its synthetic utilization is limited due to its intrinsic instability towards facile oxidative decomposition. Inspired by recent works on the reduction of ketone-substituted aza-triangulene by using atomic hydrogen to restore magnetism on surface[34,35], oxidized olympicene (olympicenon) is chosen as a precursor for on-surface Ullman coupling. This choice is made because olympicenon is stable and allows for convenient synthesis in solution, while also minimizing the side products resulting from random radical fusion during Ullmann coupling on the surface. The desired spin-1/2 Heisenberg chain can be finally obtained by on-surface reduction of the ketone groups using atomic hydrogen and subsequent tip-induced extra hydrogen desorption.

As shown in Fig. 1a, the precursor **1** was synthesized through traditional solution chemistry (Fig. S1 for solution synthesis and characterization), which is composed of a 6H-benzo[cd]pyren-6-one (olympicenone) core equipped with two Br groups. Precursor **1** was sublimated onto Au(111) and annealed at 543K to trigger Ullman coupling reactions. Afterward, the sample was exposed to atomic hydrogen, resulting in the formation of olympicene oligomeric chains with varying number of units and lengths (Fig. 1b). To characterize the chemical structure of product in each step, bond-resolved nc-AFM was carried out using CO-functionalized tip[36]. The nc-AFM image in Fig. 1c reveals the molecular structure of olympicenone chains after Ullman coupling, where each olympicenone unit displays four resolved benzene rings, with a weak contrast showing a dark appearance on the ring with a carbonyl group. This observation is consistent with similar findings for benzo[cd]pyrene[37,38]. In olympicenone chains, the π-electron magnetism is entirely quenched by the presence of carbonyl groups, as such no spin features could be detected (Fig. S10). When olympicenon oligomers are exposed to atomic hydrogen, the ketone groups are reduced. As shown in Fig. 1d, the oxygen atoms are successfully removed, leading to the generation of a $sp^3$ methylene group at the substitution site. This gives rise to a bright contrast at the $sp^3$ carbon site due to additional hydrogen association[39,40]. In the final step, the additional hydrogen atom at the $sp^3$ carbon site can be detached by applying a controllable pulse voltage (atom manipulation), reintroducing of π-electrons to the system. Following this stepwise procedure, the olympicene spin-1/2 Heisenberg chain can be constructed with different number of olympicene units and lengths.

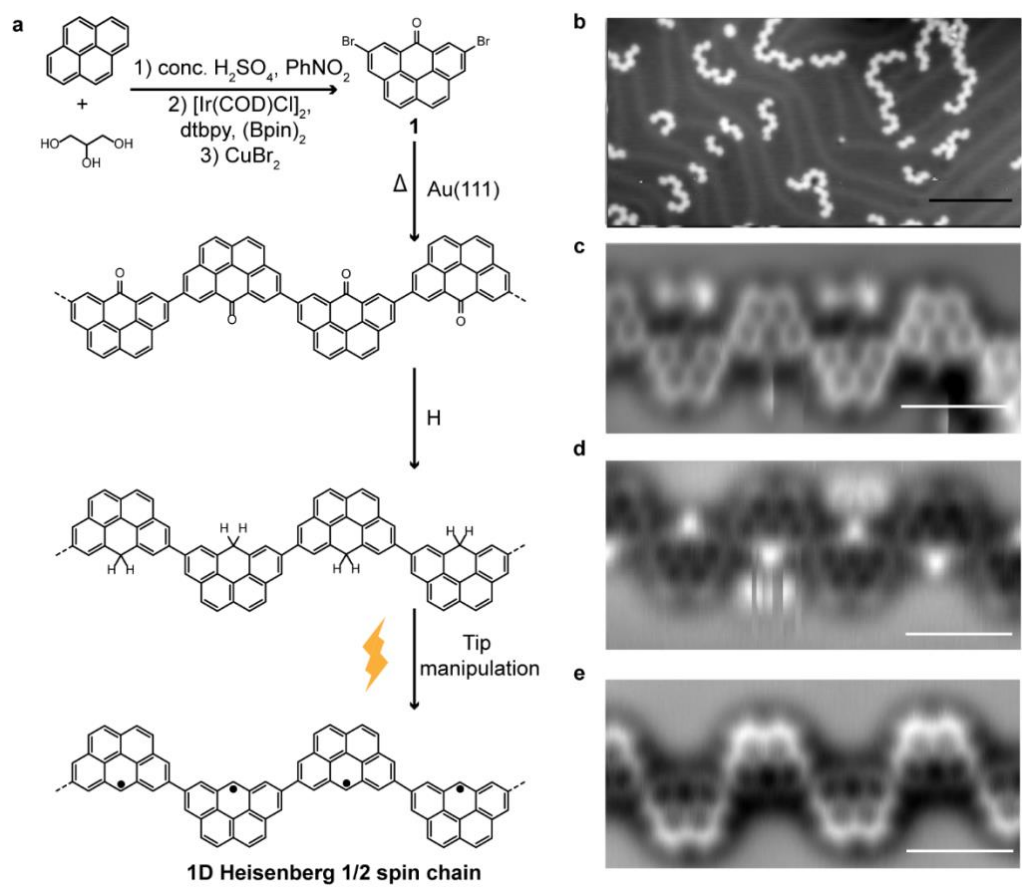

**Figure 1. Spin-1/2 Heisenberg chains fabricated through a stepwise solution/on-surface synthetic process.** (a) An overall synthetic scheme to the 1D spin-1/2 Heisenberg chains. (b) Large-scale constant-current STM image ($V$ = 300 mV, $I$ = 50 pA) of olympicene Heisenberg chains with different numbers of olympicene unit. (c-e) nc-AFM images of reaction products after Ullman coupling: (c) before and (d) after atomic hydrogen reduction and (e) upon sequential atom manipulation treatments. Scale bars: 10 nm (b); 1nm (c).

**Spin excitations in spin-1/2 Heisenberg chains of variable lengths**

The long olympicene spin chains (OSCs) primarily consist of mixed *cis-/trans*-coupling configurations with olympicene building blocks (see Fig. 2a). The d*I*/d*V* spectra of the dimer (shown in Fig. 2b) revealed that conductance steps are both below and above the Fermi level in a symmetrical manner. These steps correspond to the spin excitation from the singlet ground state to the triplet excited state. The excitation energy, precisely determined by the second derivative of d*I*/d*V* spectroscopy, signifies the energy difference between the ground and first excited states. Our MFH calculations also confirmed that two 1/2 spins of nearest-neighboring olympicene units are antiferromagnetically coupled (Fig. S11). Notably, for the olympicene dimer, its excitation energy representing the exchange interaction between the nearest neighbors is about 38 meV and 37 meV for the *cis-* and *trans*-configuration, respectively (Fig. S12). The exchange interaction remains almost identical for *cis-/trans*-coupling, allowing for a systematic investigation of the spin excitation of OSCs with varying lengths.

In Fig. 2b, d*I*/d*V* spectroscopy was conducted on OSCs with 2, 3, 4, 5, 6, and 7 units in a low-bias regime. The even-odd parity effect was observed, revealing a distinct spin excitation features in the even- and odd-numbered chains. Only conductance steps symmetrically below and above the Fermi level were detected for even-numbered OSCs, corresponding to the spin excitations between a magnetic ground state and excited states in OSCs. For odd-numbered OSCs, in addition to spin excitations, zero energy peaks were observed at the odd sites, corresponding to the Kondo resonances due to the residual 1/2 spin. The spin excitation spectra can be reproduced by simulation using the tip-surface tunneling as the scattering process up to second order[41]. The even-odd parity effect is attributed to the different ground states of even- and odd-numbered OSCs. To illustrate the experimental observations, the 1D Heisenberg model was utilized, where each olympicene unit is described as S = 1/2 with the nearest-neighbor antiferromagnetic exchange interactions of 38 meV. According to Heisenberg model calculations (Fig.3a), the ground state of short even-numbered chains is the singlet state, while the ground state of short odd-numbered chains is the S = 1/2 doublet state. The first excitation energy of even-number chains decreases from 38 meV to 7.7 meV once the chain length increases from 2 to 16 units (Fig.3b), agreeing well with the Heisenberg model calculation results (Fig.S13). For the odd-numbered OSCs, the first excitation energy decreases from 36.5 meV to 16.6 meV as the chain length increases from 3 units to 9 units, also consistent with the Heisenberg model calculations (Fig.S13). The spin excitation gaps of both even- and odd-numbered chains display a power-law decay with respect to chain lengths, which also implies a gapless excitation when the OSC reaches a threshold of olympicene units. The d$I$/d$V$ spectra of OSCs with other unit numbers are shown in Fig. S14.

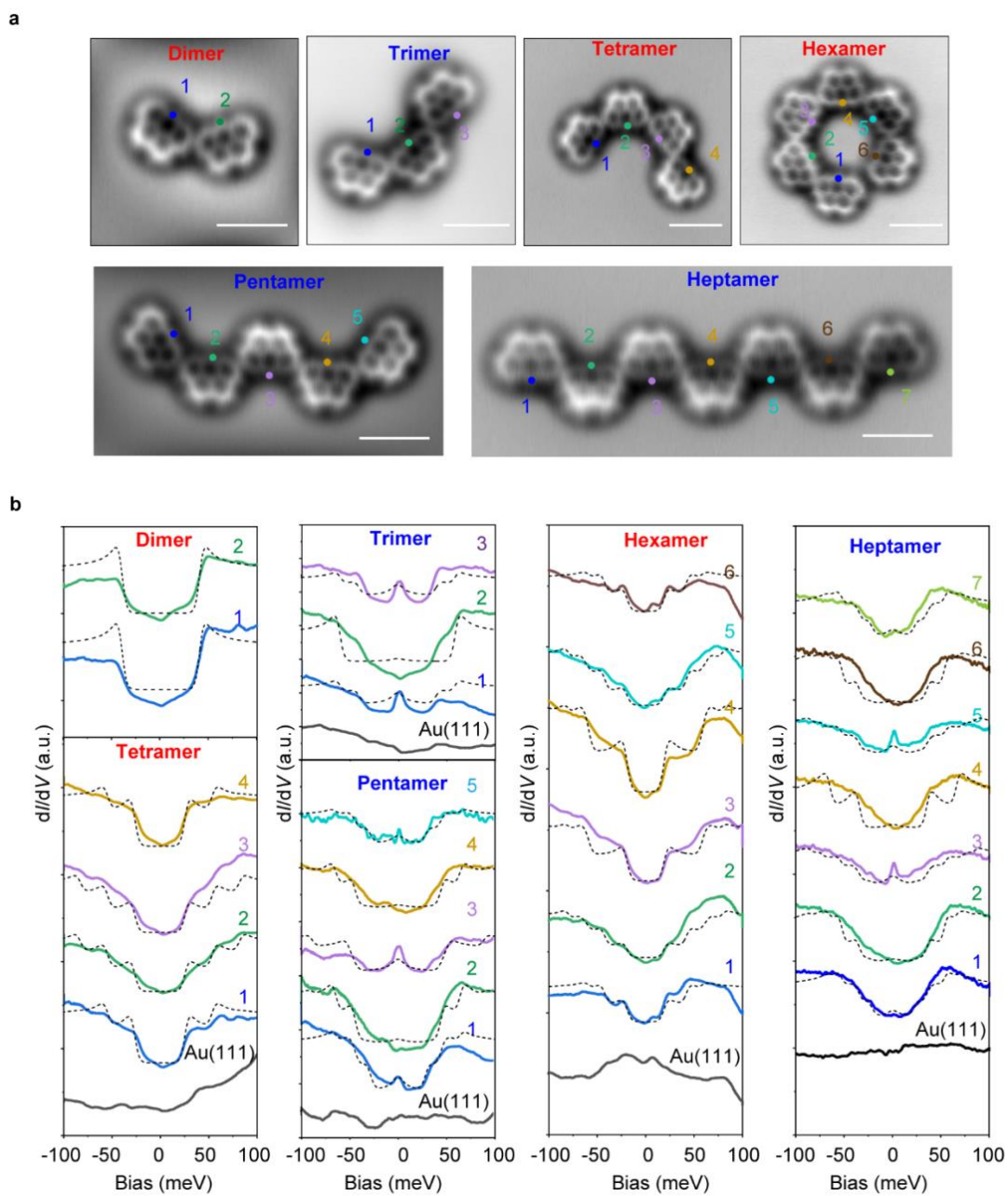

**Figure 2. d*I*/d*V* spectra measured on OSCs with different number of olympicene units.** (a) nc-AFM images of OSCs with 2, 3, 4, 5, 6 and 7 olympicene units, and (b) their corresponding experimental and simulation d*I*/d*V* spectra plotted in solid and dotted lines respectively. Setpoint: $V = 100$ mV, $I = 500$ pA, $V_{rms} = 2$ mV. Scale bars: 1 nm.

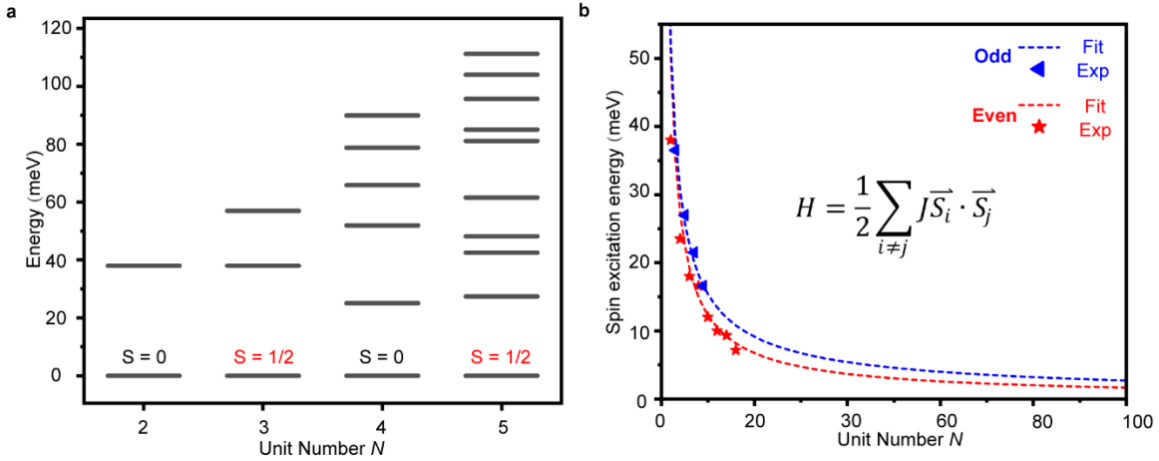

**Figure 3. Calculated ground state and spin excitations of OSCs with different number of olympicene units.** (a) Calculated low-lying spin states of OSCs with 2 to 5 olympicene units by Heisenberg model. (b) Experimental energy difference of OSCs between the ground state and first excited state as a function of olympicene units fitted by the curve of power-law function.

**Spinon dispersion obtained at nanometer scale**

The elementary excitations of spin-1/2 Heisenberg chain are the fractional quasiparticle excitations called spinons, which are created and detected in pairs. The dispersion of the lowest-lying excited states is sinusoidal, given by $\frac{\pi}{2}J|\sin(\vec{a} \cdot \vec{q})|$, where $q$ represents the wave vector along the chain length and $a$ is the lattice spacing between units[42]. Although spinon dispersion has been detected by neutron scattering spectroscopy, it has not been detected at atomic scale due to the lack of suitable platforms for high lateral resolution experimental techniques. Therefore, it is interesting to explore whether spinon dispersion can be probed in our fabricated OSCs. Spinons are usually generated in a 1D Heisenberg chain by an elementary spin-flip process, which can be realized by inelastic scanning tunneling spectroscopy measurements. When the tunneling electrons have enough kinetic energy to excite a spin-flip, a step will be generated in the differential conductivity d$I$/d$V$ or a peak in the second derivative tunneling spectra d$^2I$/d$V^2$. In one dimension, if the spinons are created, they will be scattered back and forth between the two interface boundaries, thus leading to spinon standing waves with quantized momentum determined by $\frac{2\pi n}{L}$. The amplitude of the spinon standing wave is proportional to the spatially-resolved d$^2I$/d$V^2$ in the experiment. To extract spinon dispersion information in momentum space, the d$^2I$/d$V^2$ spectra should be Fourier transformed (FT-d$^2I$/d$V^2$). Consequently, in order to obtain a better dispersion relation, a long OSC with 16 olympicene units was investigated, whose STM and nc-AFM images are shown in Fig. 4a. The inelastic scanning tunneling spectra were taken at the same position of each unit (Fig. 4b). As illustrated in Fig. 4c, FT-d$^2I$/d$V^2$ exhibits a continuum with two prominent sinusoidal features, where the lowest excitation gap locates at the reciprocal point of 0, $\pi$ and $2\pi$, and the maximum excitation energy locates at the reciprocal point of $\frac{\pi}{2}$ and $\frac{3\pi}{2}$. The result is fully consistent with the dispersion relation of $\frac{\pi}{2}J|\sin(\vec{a}\cdot\vec{q})|$. Although spin excitation steps can be observed at both positive and negative biases, the dispersion relation obtained at the negative bias is not obvious as that of the positive bias branch due to particle-hole asymmetry. To elucidate the nature of these low-lying excitations in OSCs with 16 olympicene units, we further performed exact diagonalization calculations using the Heisenberg model. As shown in Fig. 4d, the calculated spinon excitation spectra agree well with the experimental observations, suggesting that the Heisenberg model is able to capture both the ground magnetic states and the low-energy magnetic excitation dispersions.

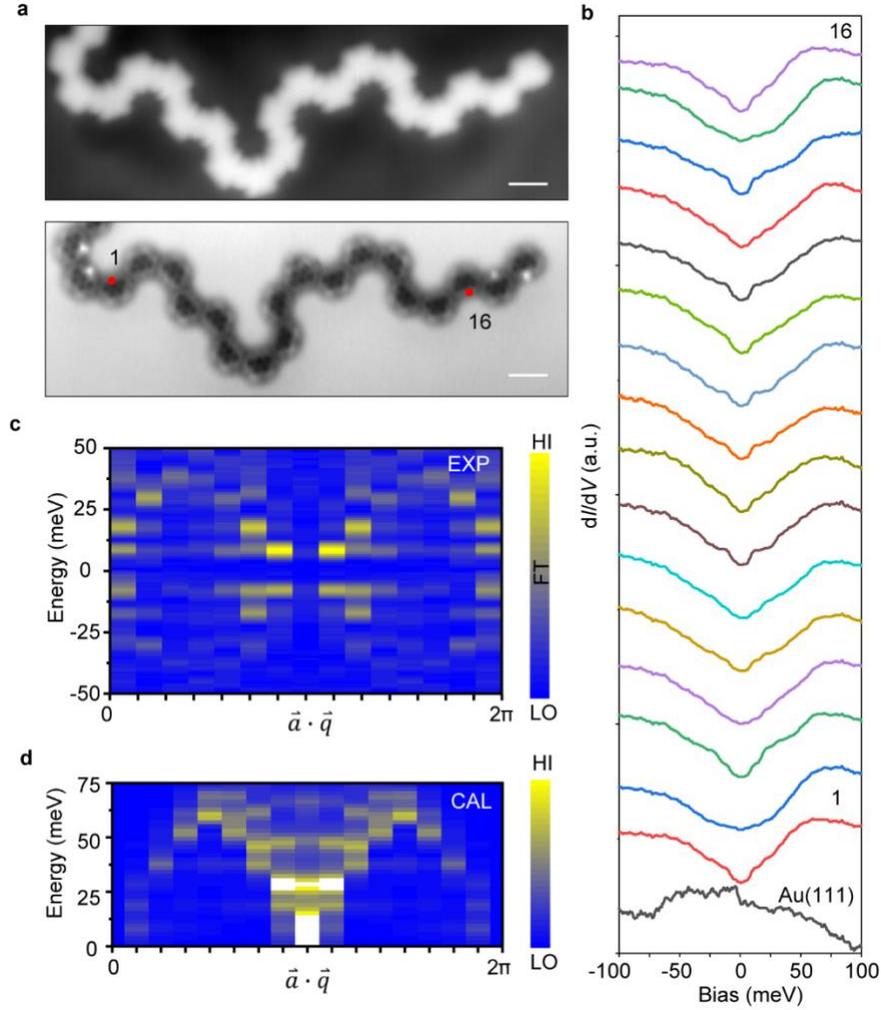

**Figure 4. Spinon dispersion in OSCs.** (a) STM ($V$ = 100 mV, $I$ = 50 pA) and nc-AFM images of OSCs with 16 olympicene units (scale bar: 1nm) and its corresponding (b) d$I$/d$V$ spectra and (c) FT- d$^2I$/d$V^2$ map ($V$ = 100 mV, $I$ = 3nA, $V_{rms}$ = 2 mV). (d) Calculated spinon dispersion of OSCs with 16 units by using Heisenberg model.

**Conclusions**

In summary, we report in this work a precise fabrication of spin-1/2 Heisenberg chains by a tandem on-surface synthesis and reduction approach. The chemical structure and collective quantum magnetism are characterized by a combination of nc-AFM and STS measurements. The even-odd parity effect is observed for short spin-1/2 chains due to the different magnetic ground states between short even-and odd-numbered OSCs. The first excitation gap of both even- and odd-numbered OSCs show power-law decay with the chain lengths[43], indicating a gapless state at long lengths. More interestingly, fractional spinon excitation is experimentally detected on long spin-1/2 chains, displaying a sinusoidal dispersion consistent with the Heisenberg calculations. Our results demonstrate a new synthesis strategy for realizing quantum magnetic materials through combined on-surface synthesis and reduction, thereby holding great promise for future fabrication and exploration of other exotic quantum materials.

## Methods

**Sample preparation and STM/AFM measurements.** The STM/AFM experiments for the electronic and chemical structure characterization were performed at 4.7 K with a commercial Createc LT-STM/qplus AFM. The Au(111) single-crystal was cleaned by cycles of argon ion sputtering and subsequently annealed to 800 K to obtain atomically flat terraces. Molecular precursors **1** were thermally deposited on a clean Au(111) surface, and subsequently annealed to 543K to fabricate olympicenone oligomers. For H-treatment, the sample was kept at 473K and was exposed for 15mins at H atoms atmosphere. The H atoms were produced by a commercial FerMi H-source at $5 \times 10^{-7}$ mbar $H_2$ atmosphere by heating with an e-beam (30W). The AFM measurements were performed with a qPlus sensor with a resonance frequency of 32.6 KHz and an oscillation amplitude of 50 pm. d$I$/d$V$ measurements were performed with an internal lock-in amplifier at a frequency of 273 Hz. Lock-in modulation voltages for individual measurements were provided in the respective figure caption. All STM/STS and AFM measurements were carried out with a CO-functionalized tungsten tip. To obtain constant-height AFM images, the tip-distance was adjusted to a few hundred pm from the STM set point $V = 300$ mV, $I = 50$ pA.

**Calculation of the OSCs.** We employed exact diagonalization to calculate the ground state and spectrum of the $S = 1/2$ Heisenberg chain with length $L$, given by

$$H = \sum_{i=1}^{L-1} S_i \cdot S_{i+1}.$$

To establish a connection with the spinons dispersion observed in the experiment, we calculate the time-dependent spin-spin correlation function:

$$\langle S_j^z(t) S_{L/2}^z(0) \rangle = \langle 0 | U^\dagger(t) S_j^z U(t) S_{L/2}^z | 0 \rangle$$

based on the ground state $|0\rangle$, where $U(t) = e^{-itH}$ represents the time-evolution operator. The dynamic spin structure factor is then obtained through the double Fourier transform of the correlation function:

$$S^{zz}(q,\omega) = \frac{1}{L} \sum_{j=1}^{L} e^{-iq(j-L/2)} \int_{-\infty}^{\infty} dt \, e^{i\omega t} \langle S_j^z(t) S_{L/2}^z(0) \rangle \cong$$

$$\frac{2\pi}{LT} \delta \sum_{j=1}^{L} e^{-iq(j-L/2)} 2 \sum_{n=0}^{N} e^{i(\omega+i\eta)t_n} Re \langle S_j^z(t_n) S_{L/2}^z(0) \rangle.$$

Here, we truncate the integral at a finite evolution time $T = 50$ and discretized the time evolution with a time interval $\delta = 0.1$. Additionally, to mitigate spectral leakage caused by the finite evolution time, we introduce a dumping factor $\eta = 0.05$. As shown in the main text, the obtained spin spectrum $S^{zz}(q,\omega)$ agrees well with the experimental observations on $L = 16$ chain.

## Data availability

All data generated in this study are available within the article and supplementary information, or from the corresponding authors upon request. Source data are provided with this paper.

## Acknowledgements

P.Y. gratefully acknowledges the financial support from the Science and Technology Commission of Shanghai Municipality (20ZR1436900) and ShanghaiTech start-up funding. X.S. acknowledges the Postdoctoral Science Foundation of China (2021M702188). C.L. acknowledges the NSFC grant 12304230 and Postdoctoral Science Foundation of China (grant GZB20230422). Y.F.J. acknowledges support from the National Program on Key Research Project under Grant No.2022YFA1402703.

## Competing interests

The authors declare no conflict of interest.

## Additional information

**Correspondence** and requests for materials should be addressed to K.Y. (email:yankk@shanghaitech.edu.cn), C.L. (email: lic_18@sjtu.edu.cn), Y.J. (email: jiangyf2@shanghaitech.edu.cn), P.Y. (email: yuping@shanghaitech.edu.cn).